\documentclass[12pt,preprint,superscriptaddress,notitlepage]{revtex4-1}
\usepackage{graphicx}  
\usepackage{dcolumn}   
\usepackage{bm}        
\usepackage{amssymb}   
\usepackage{amsmath}   
\usepackage{amsthm}    
\usepackage{xcolor}    
\usepackage{tikz}      
\usepackage[normalem]{ulem}
\usepackage{ifthen}    
\usepackage{multirow}  
\usepackage{tabu}      
\usepackage{subfigure}
\usepackage{bbold}
\usepackage{xspace}
\usepackage{siunitx} 
\usepackage{natbib}
 \usepackage[bottom]{footmisc} 
\usepackage{etoolbox}
\patchcmd{\section}
  {\centering}
  {\raggedright}
  {}
  {}
\patchcmd{\subsection}
  {\centering}
  {\raggedright}
  {}
  {}
\usepackage[utf8]{inputenc}
\usepackage[unicode=true,
  linktocpage,
  linkbordercolor={0.5 0.5 1},
  citebordercolor={0.5 1 0.5},
  linkcolor=blue]{hyperref}
\newcounter{para}

\begin{document}
	
	
\title{Gate-tunable anomalous Hall effect in a 3D topological insulator/2D magnet van der Waals heterostructure}


\author{Vishakha Gupta
}
 \thanks{denotes equal contribution}
\author{Rakshit Jain
}
 \thanks{denotes equal contribution}
\affiliation{Cornell University, Ithaca, NY 14850, USA}
\author{Yafei Ren}
\affiliation{Department of Materials Science and Engineering, University of Washington, Seattle, Washington 98195, USA}
\author{Xiyue S. Zhang}
\affiliation{Cornell University, Ithaca, NY 14850, USA}

\author{Husain F. Alnaser}
\affiliation{Department of Materials Science and Engineering, University of Utah, Salt Lake City, Utah
84112, USA}

\author{Amit Vashist}
\affiliation{Department of Materials Science and Engineering, University of Utah, Salt Lake City, Utah
84112, USA}
\affiliation{Department of Physics and Astronomy, University of Utah, Salt Lake City, Utah 84112, USA}
\author{Vikram V. Deshpande}
\affiliation{Department of Physics and Astronomy, University of Utah, Salt Lake City, Utah 84112, USA}

\author{David A. Muller}
\affiliation{Cornell University, Ithaca, NY 14850, USA}
\affiliation{Kavli Institute at Cornell, Ithaca, NY 14853, USA}

\author{Di Xiao}
\affiliation{Department of Materials Science and Engineering, University of Washington, Seattle, Washington 98195, USA}
\affiliation{Department of Physics, University of Washington, Seattle, Washington 98195, USA}

\author{Taylor D. Sparks}
\affiliation{Department of Materials Science and Engineering, University of Utah, Salt Lake City, Utah
84112, USA}

\author{Daniel C. Ralph}
\affiliation{Cornell University, Ithaca, NY 14850, USA}
\affiliation{Kavli Institute at Cornell, Ithaca, NY 14853, USA}
\date{\today}
\maketitle
\section{Abstract}
 We demonstrate advantages of samples made by mechanical stacking of exfoliated
van der Waals materials for controlling the topological surface state of a 3-dimensional
topological insulator (TI) via interaction with an adjacent magnet layer.  We assemble bilayers with pristine interfaces using exfoliated flakes of the TI BiSbTeSe$_2$ and the magnet Cr$_2$Ge$_2$Te$_6$, thereby avoiding problems caused by interdiffusion that can affect interfaces made by top-down deposition methods.  The samples exhibit an anomalous Hall effect (AHE) with abrupt hysteretic switching.
For the first time in samples composed of a TI and a separate ferromagnetic layer, we
demonstrate that the amplitude of the AHE can be tuned via gate voltage with a strong
peak near the Dirac point. This is the signature expected for the AHE due to Berry curvature
associated with an exchange gap induced by interaction between the topological surface state
and an out-of-plane-oriented magnet.


\section{Background on Topological Insulator/Magnet Heterostructures}

Interactions between three-dimension topological insulators (TIs) and magnets can induce exotic topological phases like the quantum anomalous Hall~\cite{yu2010quantized, chang2013experimental,deng2020quantum,deng2021high}  or axion insulator states~\cite{mogi2017magnetic, liu2020robust}, and might be used to harness the properties of topological-insulator surface states in spintronic devices~\cite{tokura2019magnetic}.  Coupling of a topological-insulator surface state to a magnetization that is perpendicular to the surface is predicted to open an exchange gap at the Dirac point, thereby inducing Berry curvature in the electronic states near this gap~\cite{yu2010quantized, chang2013experimental}.  If the electron chemical potential is swept through the states in the vicinity of the exchange gap, this Berry curvature should cause an anomalous Hall effect (AHE) peaked when the chemical potential is in the gap.  For an ideal scenario at low temperatures, the peak value of the Hall conductivity for an exchange gap on just one surface of a TI thin film should reach a quantized value, $e^2/2h$~\cite{xiao2010berry,mogi2022experimental}.  Previous experiments have explored coupling of magnetism to TIs via two strategies: (1) by making the TI itself magnetic (by introducing magnetic dopants~\cite{chang2013experimental, hor2010development, chang2015high, he2018topological} 
or by using an intrinsically magnetic TI material~\cite{ deng2020quantum, liu2020robust}) in which case the maximum temperature to which magnetism persists is generally less than about 40 K~\cite{mogi2017magnetic}, 
or (2) by proximity coupling of a non-magnetic TI deposited onto a separate magnetic layer, or vice-versa (using such magnets as EuS~\cite{Wei2013, Katmis2016}, 
Y$_3$Fe$_5$O$_{12}$~\cite{lang2014proximity, jiang2015independent,liu2020changes}, BaFe$_{12}$O$_{19}$~\cite{li2019magnetization}, 
Cr$_2$Ge$_2$Te$_6$~\cite{Alegria2014, Chong2018, yao2019record, Mogi2019, Mogi2021}, 
Tm$_3$Fe$_5$O$_{12}$~\cite{tang2017above, chen2019topological}, 
Zn$_{1-x}$Cr$_x$Te \cite{watanabe2019quantum}, Ga$_{1-x}$Mn$_{x}$As~\cite{Lee2018},
 Eu$_3$Fe$_5$O$_{12}$~\cite{zou2022enormous},  or MgAl$_{0.5}$Fe$_{1.5}$O$_4$ \cite{Riddiford2022})  
in which case for many materials the magnetism survives to much higher temperatures.  
However, the second strategy -- implementing well-controlled proximity-coupled TI/magnet samples -- has proved to be challenging, in that the resulting AHE signals are often very small and occasionally even non-hysteretic, so that in some cases it can be difficult to distinguish an AHE signal due to an exchange gap from nonlinear Hall responses expected for two-band conduction.
References~\cite{bhattacharyya2021recent, grutter2021magnetic}
present excellent recent reviews. 
As of yet no proximity-coupled TI/magnet samples have been fabricated with sufficient control to exhibit the full prototypical response expected from the creation of an exchange-gap – observation of a peak in the AHE as a function of gating the electron chemical potential through the gap.  Materials issues that have made such measurements challenging include bulk conduction in the TI, spatial inhomogeneities due to dopants, difficulty in positioning the energy of the Dirac point to be well-separated from bulk bands, and  disorder at the interface.

Here, we study TI/magnet bilayers using an all-van der Waals (vdW) heterostructure, by stacking together in a glovebox exfoliated flakes of the TI BiSbTeSe$_{2}$ and the insulating magnet Cr$_2$Ge$_2$Te$_6$.  In contrast to previous work on TI/magnet samples grown by top-down deposition methods, or made by assembling exfoliated flakes with air exposure~\cite{Chong2018}
, the use of a glovebox-assembled vdW structure ensures a defect- and diffusion-free atomically-flat interface~\cite{geim2013van}. Further, our use of a thin insulating vdW magnet enables both spatially-uniform magnetic coupling to the topological surface state and large tunability of the electron chemical potential using electrostatic gating.  We measure the anomalous Hall response while continuously controlling the contribution of the surface state by gating. The exfoliated samples exhibit an abrupt hysteretic switching signal for measuring the AHE, and for the first time in any TI/magnet bilayer structure we observe a clear peak in the amplitude of the AHE when the chemical potential is gated through the 
Dirac point.
This gate dependences is the signature expected from Berry curvature induced by an exchange gap, as distinct from
other possible mechanisms such as a non-topological magnetized conducting layer at the interface~\cite{huang2012transport}
or spin Hall magnetoresistance~\cite{nakayama2013spin}.  
Our interpretation is corroborated by theoretical calculations of the AHE produced by an exchange gap. 
Our results demonstrate the advantages of a clean all-vdW exfoliated topological insulator-2D magnet system for manipulating topological surface states, and pave the way for improved control of topological magneto-electric effects in TI/magnet heterostructures.

\section{Materials Characterization and Sample Assembly}

In the most-commonly-studied 3D TIs (e.g., Bi$_2$Se$_3$, Bi$_2$Te$_3$), defects and unintentional doping can result in significant bulk conduction, making it difficult to isolate or tune the properties of the surface states. In contrast, for quarternary TIs like Bi$_{2-x}$Sb$_x$Te$_{3-y}$Se$_y$, it has been shown that by optimally tuning the Sb and Se compositions the surface Dirac cone can be engineered to lie isolated inside a well-formed bulk band gap and close to the Fermi energy \cite{Arakane2012,Xu2014, Han2018}.  For this reason, we utilize exfoliated films of quaternary BiSbTeSe$_2$ (BSTS) in our samples, similar to those shown to be of high quality in ref.~\cite{Han2018}. 
In Fig.\ 1c, we show the measured longitudinal resistance $R_{xx}$  for an exfoliated 12 nm flake of BSTS (with no magnetic layer) as a function of bottom gate voltage at 2 K. The surface can be gated through the Dirac point as evidenced by the peak in the resistance and change in sign of the Hall coefficient ($R_H$) around $V_{bg} \sim -15$ V. The inset in Fig.\ 1c shows the temperature dependence of $R_{xx}$ in the same sample.  Below $\sim$ 80 K, $R_{xx}$ decreases and then saturates. This indicates that bulk conduction has largely frozen out in this regime and the transport is dominated by the 2D surfaces \cite{ ren2010large, ren2011optimizing, kushwaha2016sn, Xu2014}.

Using similar BSTS material, we fabricate dual-gated vdW heterostructure devices by stacking a few-layer flake of the vdW magnetic insulator Cr$_2$Ge$_2$Te$_6$ (CGT) on top of a BSTS flake (Fig.\ 1a). The samples are sandwiched between hBN layers (typically $<$ 10 nm) at the top and bottom to protect from degradation due to oxygen or moisture. The hBN layers also function as gate dielectrics. 
Figure 1d shows a high-angle angular dark field scanning transmission electron microscopy image of a cross section of one of our hBN/BSTS/CGT/hBN structures demonstrating clean, abrupt interfaces,
nawithout the interdiffusion that is often unavoidable in deposited heterostructures (e.g., \cite{Riddiford2022}).
A top metal electrode and a Si back gate are used to tune the carrier densities of the top and bottom surfaces of the TI, respectively. Figure 1b shows an optical image of a representative device (channel length $L=6$ $\mu$m, width $W = 2$ $\mu$m) and the configuration used for transport measurements. Platinum electrical contacts are pre-patterned on the bottom hBN layer using electron beam lithography.  

\section{Anomalous Hall Effect}

The out-of-plane magnetization of the CGT is expected to break time-reversal symmetry in the BSTS surface through proximity coupling and result in the opening of an exchange gap ($\Delta$) 
in the adjacent Dirac surface state (Fig.\ 2b), as described by the following 2D Dirac Hamiltonian: 
\begin{equation}
H(k) =   \hbar v_F(k_x\sigma_y - k_y\sigma_x) + \Delta \sigma_z  
\end{equation}
where $\sigma_{x,y,z}$ are the Pauli spin matrices, $k_x$ and $k_y$ are in-plane wave vectors, and $v_F$ is the Fermi velocity.
States in the vicinity of the gap have non-zero Berry curvature, with equal and opposite values on opposite sides of the gap. Therefore when states on opposite sides of the gap have unequal occupations, the result is a nonzero Hall conductance, $\sigma_{xy}$. The peak value of the Hall conductance, when the electron chemical potential lies in the gap, should be $e^2/2h$ in the low temperature limit~\cite{xiao2010berry,mogi2022experimental}. Consequently, a peak of the Hall resistivity ($\rho_{xy} = \sigma_{xy}/(\sigma^2_{xx}+ \sigma^2_{xy}))$, where $\sigma_{xx}$ is the longitudinal conductivity) should be found when the chemical potential is tuned through the gap. 


The panels in Fig.\ 2a and Fig.\ 2e show, respectively, the Hall voltage ($V_{xy}$) and longitudinal resistance ($R_{xx}$) measured as a function of an out-of-plane magnetic field for a dual-gated BSTS(10 nm)/CGT(2 nm) device at different fixed top-gate voltages ($V_{tg})$.  The measurements are performed at 4.4 K and under a constant bias current, with the bottom-gate voltage $V_{bg}$ fixed at 0 V. The current flows primarily through the BSTS layers since CGT is insulating at this temperature \cite{wang2018electric, lohmann2019probing, verzhbitskiy2020controlling}. The CGT flake (roughly 9 $\times$ 6 $\mu m$ in size) covers the entire device region, although because the BSTS flake (roughly 14 $\times$ 14 $\mu m$ in size) extends beyond the probed region there are additional current paths not covered by the magnetic layer that lead to shunting in $R_{xy}$. We observe hysteretic step-like changes in $V_{xy}$ corresponding to an anomalous Hall effect (AHE), indicative of a strong perpendicular anisotropy for the magnetism in CGT that is coupled to the BSTS surfaces. 
In Fig.\ 2c, we show the anomalous Hall resistance ($R_{xy}$) after subtraction of the ordinary Hall background for the measurement done at $V_{tg} = 0$ V. We find a strong remanence at zero field and coercive fields $\sim 25$ mT.
The AHE magnetic-field sweep does not contain any non-monotonicity or ``hump'' sometimes associated with a topological Hall effect.
Interestingly, the form of this AHE response in the BSTS/CGT heterostructure is quite different from the magnetic properties of standalone few-layer CGT flakes, which typically show negligible remanence at zero field and close-to-zero coercive fields \cite{Gong2017, Idzuchi2019, wang2018electric}. This difference indicates that the proximity effect at the interface of the TI/magnet enhances the perpendicular magnetic anisotropy in our samples.  

As the chemical potential of the top surface is tuned by varying the $V_{tg}$, we observe a modulation in the observed AHE signal. The amplitude of the AHE response is maximum at $V_{tg} = 0.55$ V and becomes smaller as the gate voltage is tuned on either side of this maximum. In Fig.\ 2d, we plot the extracted signal size of the anomalous Hall resistance response as a function of $V_{tg}$ (black trace). This trend tracks approximately with the gating behaviour of the longitudinal resistance $R_{xx}$
shown by the dotted blue trace. We therefore identify the maximum in the AHE response as due to tuning of the electron chemical potential within the exchange gap, as expected from the Berry curvature picture. 
The association of these changes with surface states is further supported by the observation that the slope of the ordinary Hall background in $V_{xy}$ (Fig.\ 2a) changes sign across the AHE peak (e.g. compare $V_{tg} = -1.2$ V and $V_{tg} = 2.8$ V), indicating a change in surface charge carriers from electron-like to hole-like as the Fermi level is tuned across the gap. 
Similar data from additional samples for the gating dependence of the AHE are presented in the supplemental information.

Although both the Hall resistance and the longitudinal resistance are sharply peaked, their peak values occur at slightly different values of gate voltage ($V_{tg} =$ 0.25 V for $R_{xx}$, compared to $V_{tg} =$ 0.55 V for $R_{xy}$). This difference can be explained by considering the contributions to transport from both the top and bottom surfaces of the TI. The Hall resistance should peak when the chemical potential is inside the exchange-induced band gap whereas the longitudinal resistance should peak at the chemical potential with minimal density of states, i.e., at the Dirac point of the bottom surface. The energy offset between the peaks suggest that the Dirac points of the two surfaces have a relative shift, as illustrated schematically in Fig.\ 3c where the bands are calculated based on the tight-binding Hamiltonian of Bernevig-Hughes-Zhang model with the parameters of BSTS (see details in the Supplemental Materials). 
 The existence of such a relative shift is not surprising since the bottom surface is interfaced with hBN while the top surface is coupled to the CGT layer. With a finite band gap on top and a potential energy difference between both layers, we employ the Kubo-Streda formula to calculate the resistances as a function of chemical potential as shown in Fig.~\ref{figmuR}d. The predicted trend agrees well with experiments.  Details about the calculation method are included in the supplemental material.

It would be interesting to convert the measured peak in Hall resistance to Hall conductivity $\sigma_{xy}$ to see how the peak value compares the quantized value of $e^2/2h$.  However, the fact that the large TI flake in our device extends beyond the edge of the magnetic layer causes the local resistivity to be spatially inhomogeneous, preventing a determination of $\sigma_{xy}$ within just the portion of the TI under the magnetic layer.  In future work we plan to develop fabrication techniques that will allow more-precise definition the lateral device geometry, along with tests of whether our devices can be tuned into the quantum AHE regime.

\section{Longitudinal Resistance}

We now turn to the magnetic-field dependence of the longitudinal resistance. As shown in Fig.\ 2e, we observe a trend in $R_{xx}$ versus out-of-plane magnetic field, as is expected from magneto-transport in topological insulators~\cite{wang2012room}. Interestingly, for $V_{tg}$ close to the AHE peak (i.e., as the electron chemical potental approaches the exchange gap), a butterfly pattern emerges with small peaks of enhanced conduction near the magnetic-field values where the magnetization reverses. This type of conductance enhancement has been observed previously in samples exhibiting the anomalous Hall effect \cite{Checkelsky2012, Wei2013} and has been associated with transport along magnetic domain walls present in the multidomain state near the coercive field. Although our system is likely not in the quantum anomalous Hall regime, chiral domain wall states should also exist along magnetic domain walls in our samples, leading to an analogous enhancement.  A derivation of the domain wall dispersion is included in the supplemental material. This observation provides further evidence that our measurement is probing the physics of the proximity-coupled topological states at the vdW interface.

\section{Temperature Dependence}

The temperature dependence of $R_{xy}$ for the same BSTS/CGT device as Fig.\ 2  is shown in Fig.\ 4.  The AHE response becomes smaller as the temperature is increased above the base temperature of 4.4 K and is absent above $\sim 68$ K, close to the reported $T_c$ of bulk CGT \cite{Gong2017}. This again confirms that the observed magnetism in the top surface of BSTS is induced by proximity exchange coupling with CGT. We do observe a small  enhancement in  $T_c$ for our samples compared to the reported values of few-layer CGT  (40 - 47 K for 3 - 4 layers \cite{Gong2017} and 53-57 K for 5 layers~\cite{wang2018electric}). The extracted decrease in the AHE response and in the coercive fields ($B_c$) are plotted in Fig.\ 3b.


\section{Conclusions}
We have demonstrated that the use of mechanical assembly of van der Waals materials to form a pristine interface provides a strategy that avoids materials challenges which have inhibited research progress in studying topological insulator/magnet heterostructures grown by molecular beam epitaxy or other deposition techniques. Mechanically-assembled BiSbTeSe$_2$/Cr$_2$Ge$_2$Te$_6$ bilayers exhibit a strong anomalous Hall effect (AHE) with abrupt hysteretic switching.  The amplitude of the AHE is tunable with gate voltage, with a prominent peak when the electron chemical potential is swept through the Dirac point by gating, the first time this behavior has been demonstrated in a bilayer of a topological insulator proximity-coupled to a separate magnetic layer.  This is the signature expected from the Berry curvature of the topological surface states in the presence of an exchange gap due to interaction with the magnetic layer.  The high quality of mechanically-assembled van der Waals structures provides a platform for future studies of the quantum anomalous Hall and axion insulator states, and the rich phenomenology of topological magneto-electric phenomena predicted for these states.

\section{Acknowledgements}
We thank Maciej Olszewski, Bozo Vareskic, Liguo Ma, Avi Shragai, Patrick Hollister and Thow Min Cham for assistance with sample fabrication and measurements, and Nitin Samarth for comments on a draft of the paper.
Research at both Cornell and the Univ.\ of Washington was funded by the AFOSR/MURI project 2DMagic (FA9550-19-1-0390).  Research at Cornell was also supported by the NSF (DMR-2104268) and the Cornell Center for Materials Research (CCMR, supported by the NSF via grant DMR-1719875), and utilized the shared facilities of both the CCMR and the Cornell NanoScale Facility, a member of the National Nanotechnology Coordinated Infrastructure (supported by the
NSF via grant NNCI-2025233). Research at the University of Utah was supported by the National Science Foundation under the Quantum Leap Big Idea Grant No.\ 1936383 as well as the Kuwait Foundation for the Advancement of Sciences (KFAS) under project code “CB20-68EO-01”.

\newpage

\section{References}
\bibliographystyle{naturemag}
\bibliography{ref.bib}

\begin{figure}[ht!]
    \includegraphics[width =\linewidth]{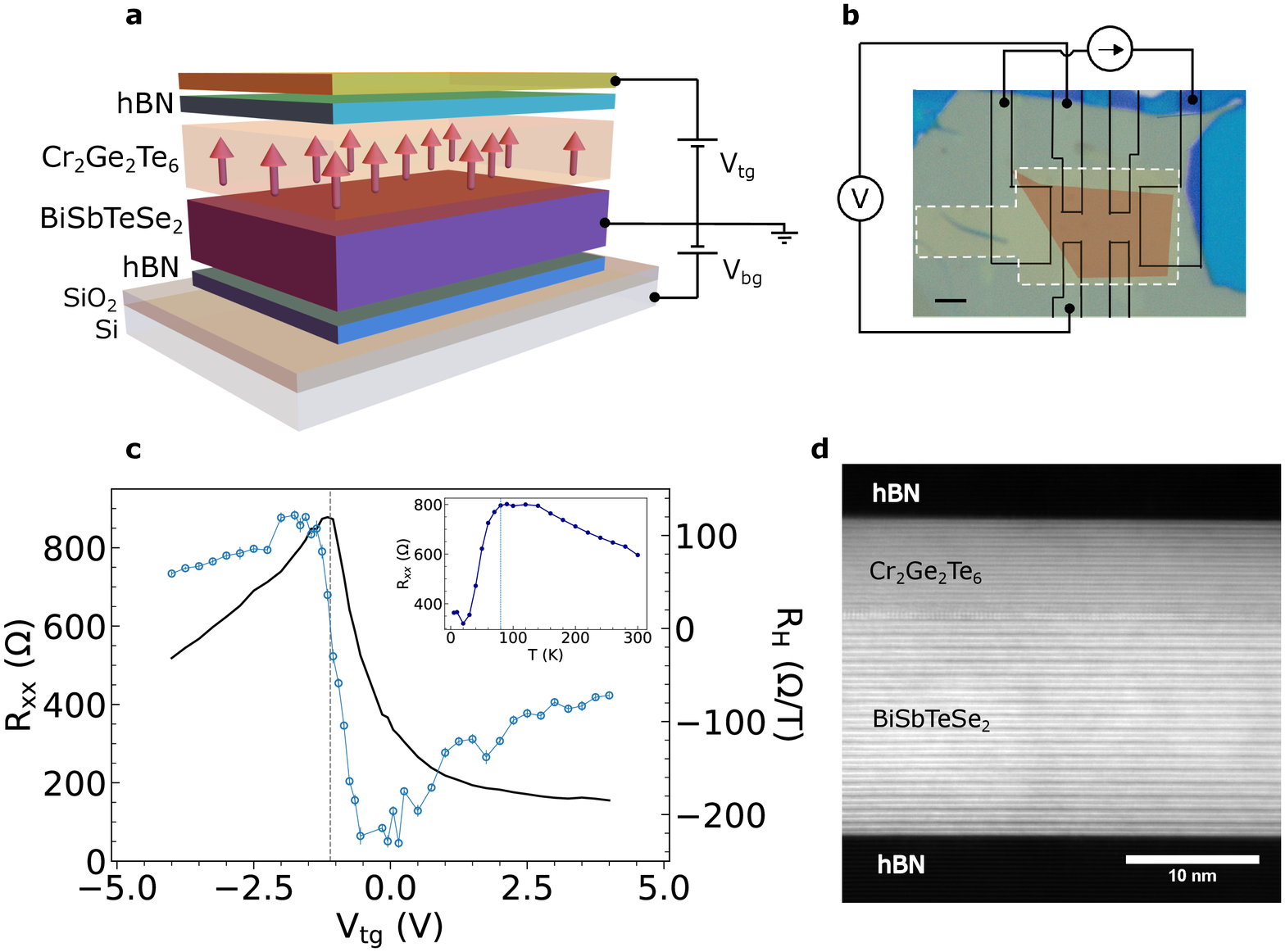}
    \caption{\textbf{Measurement setup and characterization of BiSbTeSe$_2$ flakes.} (a) Schematic of the device geometry and electrostatic gating geometry. TI and 2D magnet flakes with few-layer thickness are double-encapsulated between hBN layers on a Si/SiO$_2$ substrate. (b) Optical microscope image of a BSTS (\textit{yellow})/CGT (\textit{orange}) device with a Ti/Pt top gate (\textit{white dotted line)}, along with the measurement scheme for magneto-transport measurements. Scale bar 2 $\mu$m. (c) Longitudinal resistance (R$_{\text{xx}}$) of a 12 nm thick BSTS flake encapsulated by a top hBN (with no magnetic layer) measured as a function of bottom gate voltage  (V$_{\text{bg}}$) at $T= 2$  K and as a function of temperature for $V_{bg} = 0$ V (inset). (d) Cross-sectional HAADF-STEM image of a BSTS/CGT device.}
    
\end{figure}

\begin{figure}[ht!]
    \includegraphics[width =\linewidth]{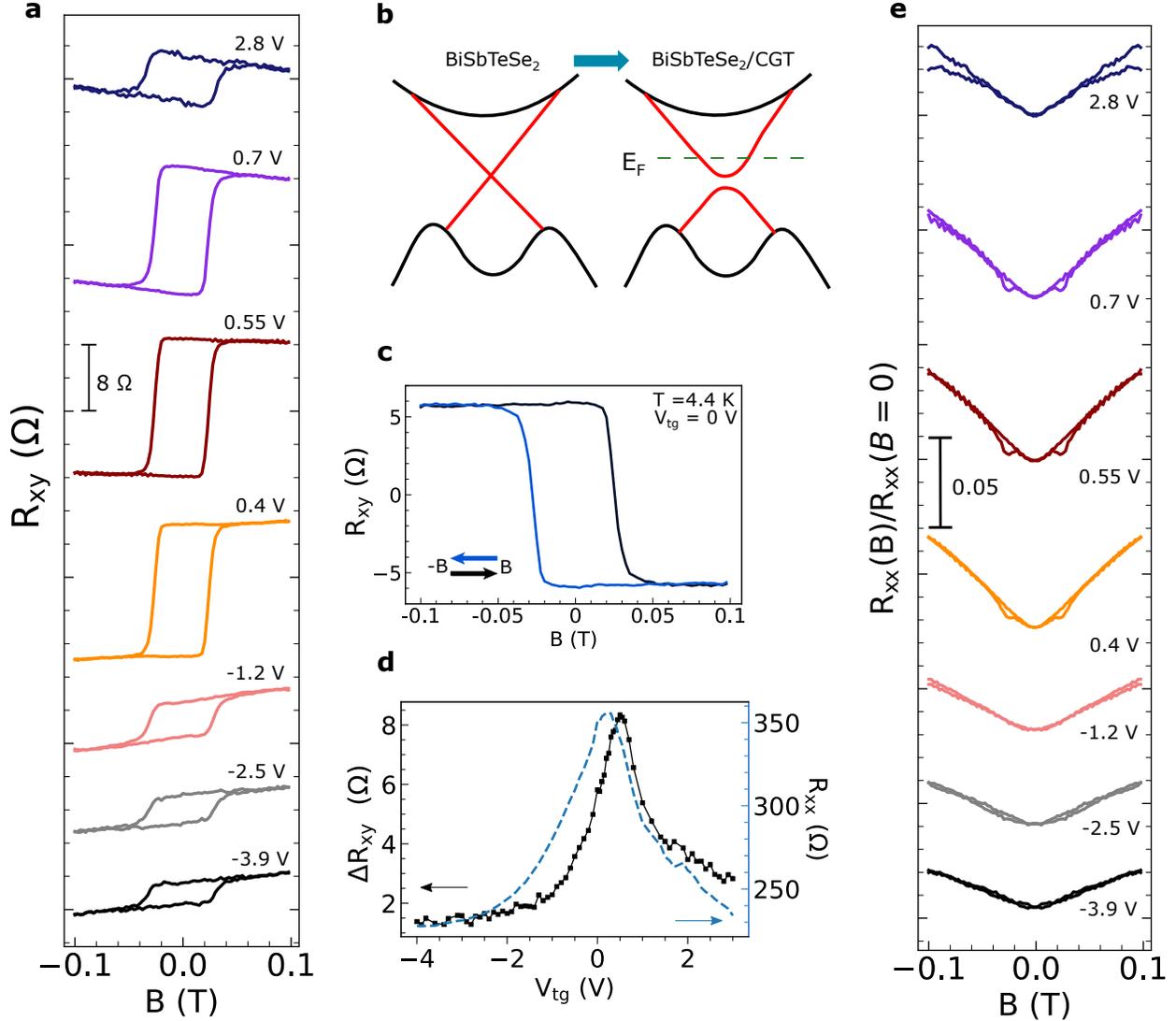}
    \caption{\textbf{Anomalous Hall effect tuned by gate voltage.} 
    (a) Hall resistance R$_{xy}$ for a BSTS/CGT (10 nm/2 nm) device measured at several top-gate voltages V$_{tg}$ at $T$ = 4.4 K. At V$_{tg} = 0.55$ the amplitude of the AHE is a maximum and the linear ordinary Hall background flips sign. (b) Schematic band diagram of a topological surface state for a TI (left) and TI/magnet (right). An exchange gap opens in the topological surface state band  due to proximity coupling with the magnet. The Fermi level (dashed line) can be tuned across this gap. (c) AHE contribution to $R_{xy}$ at V$_{tg}$ = 0 after subtraction of linear background. (d) (black solid line) Size of AHE signal $\Delta$R$_{xy}$ as a function of top-gate voltage. The trend matches closely with the observed top-gate dependence of  R$_{xx}$ (blue dashed line) measured at zero magnetic field. (e) Top-gate dependence of the longitudinal voltage V$_{xx}$ of the same device. A butterfly pattern emerges with dips in resistance near the coercive magnetic field.  These dips are  largest for voltages close to V$_{tg}$ = 0.55 V, i.e. when the Fermi level is tuned close to the magnet-induced gap in the topological surface state. }
\end{figure}

\begin{figure}
    \includegraphics[width =0.9 \linewidth]{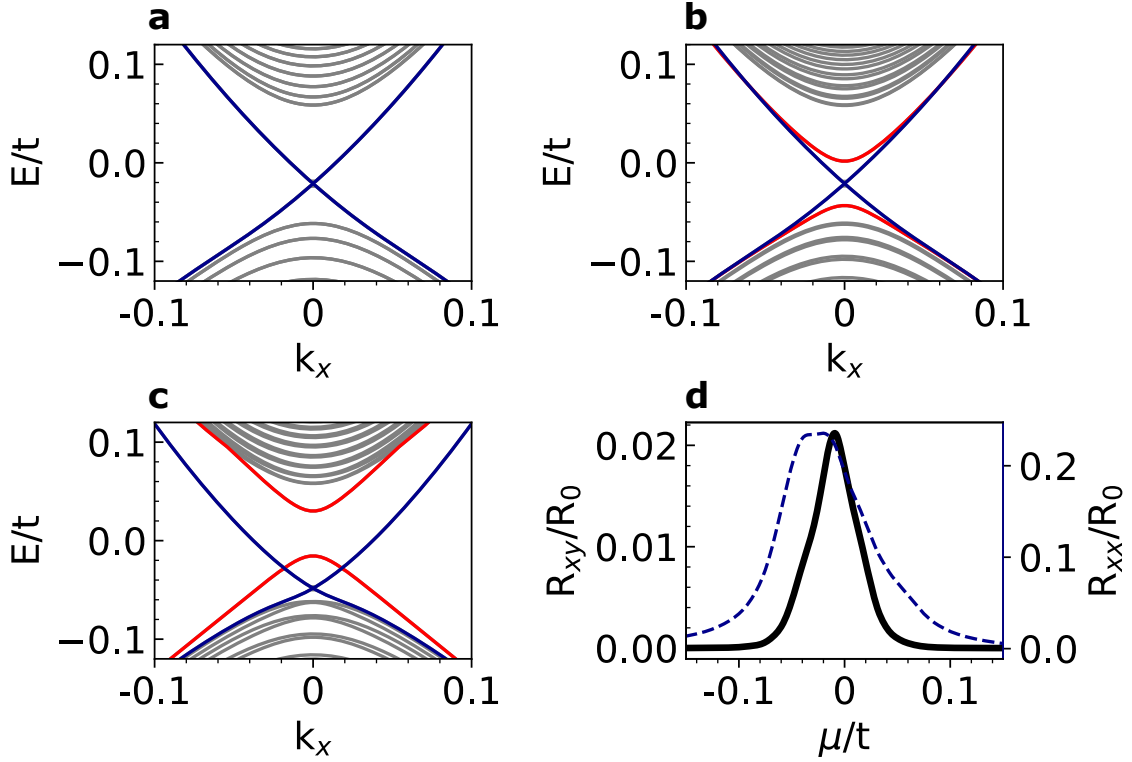}
    \caption{\textbf{Predictions for topological surface states with and without coupling to a magnetic layer.}(a) Schematic diagram of surface states for a three-dimensional topological insulator thin film with no magnetic coupling.  Red and blue curves indicate top and bottom topological surface states. (b) Surface states for a sample in which the top interface of the TI is proximately-coupled to a magnetic layer. (c) Surface with both magnetic coupling to the top TI interface and an intersurface potential difference. (d) Calculated 2D resistivity and Hall effect for a square sample (length = width) as a function of chemical potential obtained by using Kubo-Streda formula for the case corresponding to panel (c). $R_0=25.8~$k$\Omega$. }
    \label{figmuR}
\end{figure}

 \begin{figure}[ht!]
    \includegraphics[width =0.9\linewidth]{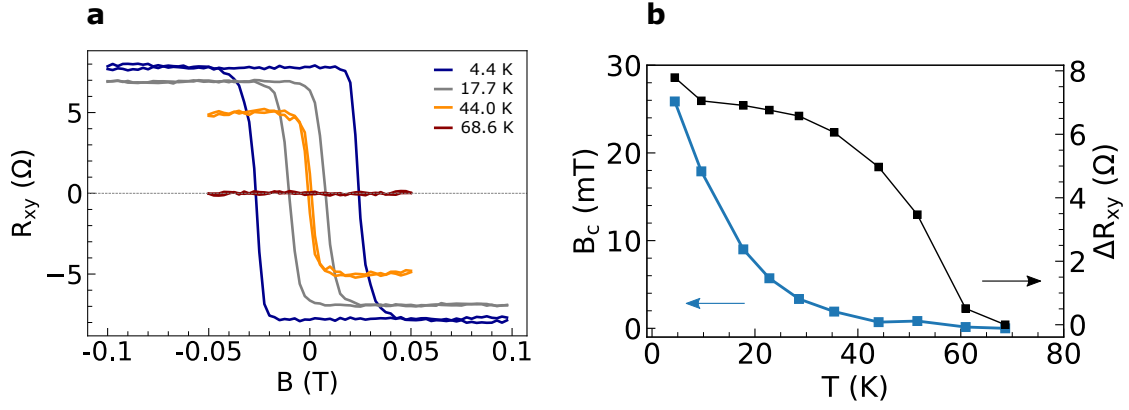}
    \caption{\textbf{Temperature dependence of magnetic properties.} (a) Extracted R$_{xy}$ vs.\ magnetic field at various sample temperatures, with the ordinary Hall contribution subtracted. (b) Temperature dependence of the coercive field (blue) and the amplitude of the AHE response (black). There is a sharp onset below 68 K. }
\end{figure}

\end{document}


	
\title{SI: Gate-tunable anomalous Hall effect in a 3D topological insulator/2D magnet van der Waals heterostructure}

\author{Vishakha Gupta}
 \thanks{denotes equal contribution}

\author{Rakshit Jain}
 \thanks{denotes equal contribution}

\affiliation{Cornell University, Ithaca, NY 14850, USA}

\author{Yafei Ren}
\affiliation{Department of Materials Science and Engineering, University of Washington, Seattle, Washington 98195, USA}

\author{Xiyue S. Zhang}
\affiliation{Cornell University, Ithaca, NY 14850, USA}

\author{Husain F. Alnaser}
\affiliation{Department of Materials Science and Engineering, University of Utah, Salt Lake City, Utah
84112, USA}

\author{Amit Vashist}
\affiliation{Department of Materials Science and Engineering, University of Utah, Salt Lake City, Utah
84112, USA}
\affiliation{Department of Physics and Astronomy, University of Utah, Salt Lake City, Utah 84112, USA}
\author{Vikram V. Deshpande}
\affiliation{Department of Physics and Astronomy, University of Utah, Salt Lake City, Utah 84112, USA}

\author{David A. Muller}
\affiliation{Cornell University, Ithaca, NY 14850, USA}
\affiliation{Kavli Institute at Cornell, Ithaca, NY 14853, USA}

\author{Di Xiao}
\affiliation{Department of Materials Science and Engineering, University of Washington, Seattle, Washington 98195, USA}
\affiliation{Department of Physics, University of Washington, Seattle, Washington 98195, USA}

\author{Taylor D. Sparks}
\affiliation{Department of Materials Science and Engineering, University of Utah, Salt Lake City, Utah
84112, USA}

\author{Daniel C. Ralph}
\affiliation{Cornell University, Ithaca, NY 14850, USA}
\affiliation{Kavli Institute at Cornell, Ithaca, NY 14853, USA}
\maketitle
\section{Supplementary information / Methods}
\section{Fabrication Details}
The dual-gated BSTS/CGT vdW heterostructure devices were fabricated using a layer-by-layer dry transfer technique described in \cite{wang2013one}, inside a glove box with O$_2$ and H$_2$O levels less than 0.5 ppm. 

Pure elemental metals Bi, Sb, Te, Se were ordered from Sigma-Aldrich Co., with a purity of 5N grade. Quartz tubes with dimensions of an outer diameter of 14 mm, an inner diameter of 12 mm and a wall thickness of 2 mm were ordered from Technical Glass Products, Inc. The raw materials were prepared with a ratio of 1:1:1:2 of Bi:Sb:Te:Se, respectively. The materials were crushed and mixed for 30 minutes producing a mixed fine powder of weight ~3 g. The powder was then loaded into a carbon-coated quartz tube. The tube is prepared for sealing by flushing with argon gas four times followed by vacuum pumping of pressure ~10–6 torr. Finally, the tube was sealed producing an ampule of length 6–8 cm. The crystal was then grown via the method described in ref. \cite{Han2018}. Bulk crystals of Cr$_2$Ge$_2$Te$_6$ and hBN were commercially obtained (HQ graphene and 2D Semiconductors, respectively). 

The individual flakes were exfoliated from the bulk crystals and picked up sequentially using a polymer stamp (Poly(Bisphenol A carbonate)). The stacked flakes were then transferred to a hBN flake which was pre-patterned with 6-nm thick Pt electrodes for making electrical contact to the BSTS and subsequently tip cleaned to remove resist residue. Finally, a metallic Ti/Pt gate was lithographically defined above the top hBN flake.    
\section{Data from additional devices}
\renewcommand{\thefigure}{S1}

 \begin{figure}[ht!]
    \includegraphics[width =\linewidth]{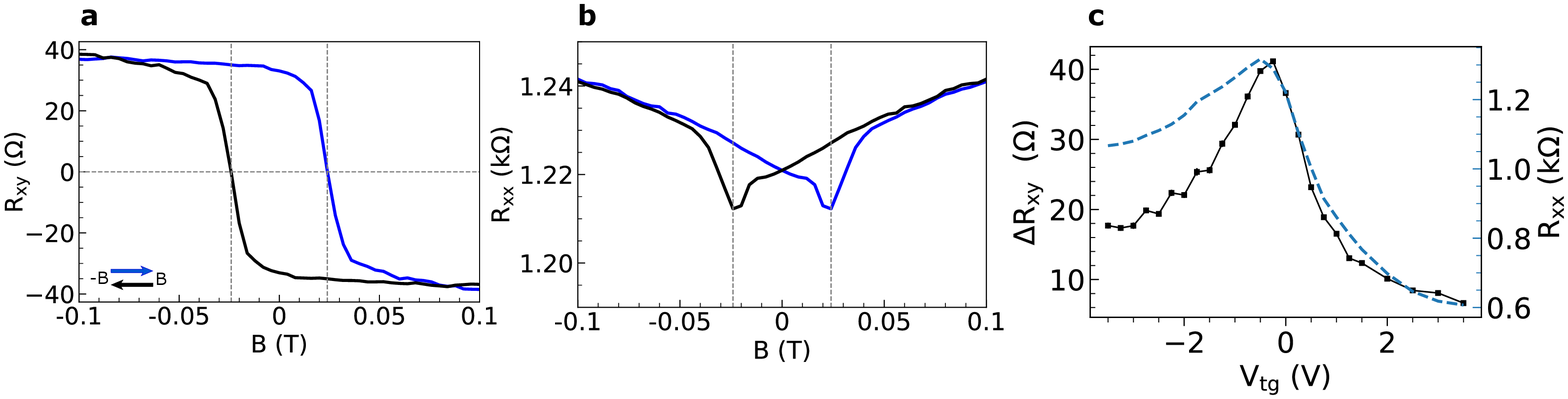}
    \caption{Additional data from a BSTS(14 nm)/CGT(4.2 nm) device (channel length L = 6 $\mu$m, width W = 2
 $\mu$m) 
  at 4.5 K. (a) Anomalous Hall resistance and (b) longitudinal resistance measured at a function of magnetic field at $V_{tg}$ = 0. (c) Top-gate dependence of the size of the AHE signal (solid black line) and $R_{xx}$ measured at zero magnetic field (dashed blue line).}
\end{figure}
\renewcommand{\thefigure}{S2}
 \begin{figure}[ht!]
    \includegraphics[width =\linewidth]{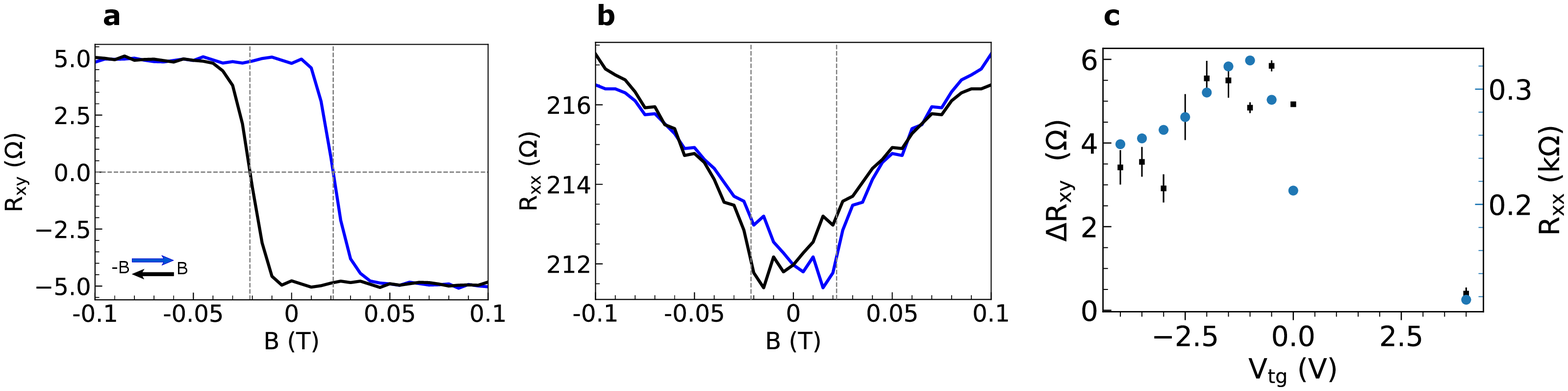}
    \caption{Additional data from a BSTS(10 nm)/CGT(5.0 nm) device (channel length L = 6 $\mu$m, width W = 2
 $\mu$m) at 4.5 K. (a) Anomalous Hall resistance and (b) longitudinal resistance measured at a function of magnetic field at $V_{tg}$ = 0. (c) Top-gate dependence of the size of the AHE signal (solid black line) and $R_{xx}$ measured at zero magnetic field (blue dashed line).}
\end{figure}

\section{Magneto-transport measurements}
Electrical transport measurements were performed in a vector magnet cryostat (1T, 1T, 7T) with a base temperature of 4.2 K for the devices studied in Fig.~2 and Fig.~3 in the main text. Characterization of BSTS flakes (Fig.~1c in the main text) was done in a Quantum Design Physical Properties Measurement System (PPMS).  A Keithley 6221 source meter was used to apply a constant ac excitation current (100 nA - 5 uA) at 97.7 Hz to the device. The longitudinal and Hall  signals were collected using lock-in techniques (Ametek DSP 7270). The top and bottom gate dc gate voltages were applied using two Keithley 2400 source meters. 
\section{Details of theoretical calculations}
\subsection{Model}
For calculating the surface state dispersion, we employ the tight-binding model Hamiltonian 
\begin{align}
\label{Ham3DTI}
    H=&M_{\bm{k}} \sigma_x + \lambda \sigma_z (s_y \sin k_x - s_x \sin k_y) + \lambda_z \sigma_y \sin k_z + M_{\bm{k}}' \sigma_0 s_0 \notag \\
    =&\begin{pmatrix}
       M_{\bm{k}}' &  -\lambda (i\sin k_x + \sin k_y) &   M_{\bm{k}}-i\lambda_z\sin k_z &  0 \\
-\lambda (-i\sin k_x + \sin k_y) &   M_{\bm{k}}' &     0    &  M_{\bm{k}}-i\lambda_z\sin k_z \\
   M_{\bm{k}}+i\lambda_z\sin k_z &         0 &   M_{\bm{k}}' &  \lambda (i\sin k_x + \sin k_y) \\
        0 & M_{\bm{k}}+i\lambda_z\sin k_z  & \lambda (-i\sin k_x + \sin k_y) & M_{\bm{k}}'  \\
\end{pmatrix}
\end{align}
where $\bm{\sigma}$ and $\bm{s}$ are Pauli matrices for orbital and spin separately. 
The mass term $M_{\bm{k}}=\Delta + 2 \sum_i t_i (1-\cos k_i a_i)$ where $a_i$ is the lattice constant along $i$-th direction with $i=x,y,z$. $\lambda$ and $\lambda_z$ are spin-orbit coupling related hopping energies. $M_{\bm{k}}'=2 \sum_i t_{0i} (1-\cos k_i a_i)$ introduces the particle-hole asymmetry. We set $a_x=a_y=a$, $t_x=t_y=t$, $t_{0x}=t_{0y}=t_0$. 
Following ref.\ \cite{https://doi.org/10.48550/arxiv.2203.14301}, we employ the parameters $t_0a^2=7.46 $ eV\AA$^2$, $t_{0z} a_z^2=-1.22$ eV\AA$^2$, $t a^2=24.7 $ eV\AA$^2$, and $t_{z} a_z^2=3.19$ eV\AA$^2$. The Fermi velocity terms are $2\lambda a = 3.23 $ eV\AA~ and $2\lambda_z a_z = 1.03$ eV\AA. The band gap is $\Delta=-0.0553 $eV. We take the lattice constant as $a=4.2 $\AA~and $a_z=1.5a$.

\subsection{Transport calculation}
We employ the Kubo-Streda formula with relaxation time approximation to calculate the conductivity \begin{align}
    \sigma_{\alpha \beta}=\frac{e^2}{h} \frac{\hbar^2}{Na^2}\sum_{\bm{k}} \int_{-\infty}^{+\infty} d\xi n_F(\xi) {\rm Tr}[\hat{v}^\alpha \frac{d G^R}{d \xi} \hat{v}^\beta (G^A-G^R) - \hat{v}^\alpha  (G^A-G^R) \hat{v}^\beta \frac{d G^A}{d \xi} ]
\end{align}
where $e$ is the elementary charge, $h$ is the Planck constant, $a$ is the lattice constant, $N$ is the number of unit cell, $n_F$ is the Fermi-Dirac distribution, $\hat{v}^\alpha=\frac{1}{\hbar}\frac{\partial H}{\partial k_\alpha}$ is the velocity operator, $G^{R/A}(\xi)=1/(\xi-H \pm i \gamma)$ are the retarded and advanced Green's functions. For simplicity, a constant self-energy $\gamma$ is employed. 

The longitudinal conductivity can be obtained as
\begin{align}
    \sigma_{\alpha \alpha} =
    \frac{e^2}{h} \frac{\hbar^2}{Na^2}\sum_{\bm{k}} \int_{-\infty}^{+\infty} d\xi \frac{1}{2} \frac{d n_F(\xi)}{d \xi} {\rm Tr}[\hat{v}^\alpha (G^R-G^A) \hat{v}^\alpha (G^R-G^A)].
\end{align}

\subsection{Domain wall state}
Here we analyze the chiral edge states along a domain wall by using a low-energy continuum model, following \cite{semenoff2008domain,ren2016topological}. In this model, the Hamiltonian reads
\begin{eqnarray}
       H= v (k_x \sigma_y - k_y \sigma_x) + \sigma_z m(x,y)
       \label{ham1}
\end{eqnarray}
where $v$ is the Fermi velocity, $\bm{\sigma}$ are Pauli matrices of spin and $m(x,y)$ represents the position dependent exchange field from the ferromagnetic order.
We consider a single domain wall along $x=0$ where the mass term from the magnetic order is positive in one side while is negative in the other side, i.e., $m(x,y)=-\Delta$ for $x>0$ and $m(x,y)=\Delta$ for $x<0$ with $\Delta$ being positive. The mass term opens an energy gap $2\Delta$ in each region. We focus on the interface and look for domain wall states with eigenenergies within the energy gap. 

To obtain these domain wall states, one can solve the Dirac equation as follows
\begin{eqnarray}
       \left[\begin{array}{cc}
              m(x,y)  &  v(k_x - ik_y)  \\
             v ( k_x + ik_y) & -m(x,y)
                \end{array}
          \right]  \left[\begin{array}{c}
              u(x,y)  \\
              v(x,y)
         \end{array}
          \right] =   E \left[\begin{array}{c}
              u(x,y)  \\
              v(x,y)
         \end{array}
          \right].
       \label{drceq1}
\end{eqnarray}
The eigenwavefunctions are expected to have the following form. Along $y$-direction, a plane wave is expected due to the transition symmetry. Along $x$ direction, localized states with an exponentially decaying wavefunction is expected since the energies are located within the energy gap. Therefore, the trial wavefunction for this equation is assumed to be, in region A,
\begin{eqnarray}
       u(x,y)=u_0 e^{-\lambda_0 x}e^{ik_y y}, \nonumber \\
	 v(x,y)=v_0 e^{-\lambda_0 x}e^{ik_y y},
       \label{wavfa}
\end{eqnarray}
in region B,
\begin{eqnarray}
       u(x,y)=u_1 e^{\lambda_1 x}e^{ik_y y}, \nonumber \\
	 v(x,y)=v_1 e^{\lambda_1 x}e^{ik_y y}
       \label{wavfcn}
\end{eqnarray}
where $\lambda_i$ ($i=0$,$1$) are positive.
By substituting this trial wavefunction into the Dirac equation, one can obtains
\begin{eqnarray}
	E^2=\Delta^2+ v^2(k_y^2-\lambda_i^2), \label{solt} \\
	\frac{v_0}{u_0}=\frac{\Delta + E}{\lambda-k_y}, \frac{v_1}{u_1}=\frac{\Delta - E}{\lambda+k_y}.
\label{solt1}
\end{eqnarray}
At $x=0$, both wavefunctions must have the same value which give rise to an additional contraint. Combining these equations, one can find that $\lambda=\Delta/v$ and the in-gap eigenstates exhibit the dispersion relation $E=-vk_y$.  Such states will therefore provide enhanced conductivity parallel to the domain wall.

\section{References}
\bibliography{ref.bib}